\documentclass[epj]{svjour}
\usepackage{graphics}
\usepackage{color}
\usepackage{graphicx}
\usepackage{dcolumn}
\usepackage{bm}
\newcommand{\U}{{\cal {U}}}
\newcommand{\dU}{d_{\cal U}}
\newcommand{\LamU}{\Lambda_{\cal U}}
\newcommand{\C}{{\cal {C}}}
\begin{document}
\title{Thermal Unparticles: A New Form of Energy Density in the Universe}
\author{Shao-Long Chen\inst{1,2} %
\and Xiao-Gang He\inst{1,2} 
\and Xue-Peng Hu\inst{3} %
\and Yi Liao\inst{3} %
\thanks{
liaoy@nankai.edu.cn}%
}                     
\offprints{}          
\institute{Department of Physics and Center for Theoretical
Sciences, National Taiwan University, Taipei %
\and Leung Center for Cosmology and Particle Astrophysics,
National Taiwan University, Taipei %
\and Department of Physics, Nankai University, Tianjin 300071}
\date{Received: date / Revised version: date}
%
\abstract{ Unparticle $\U$ with scaling dimension $d_\U$ has
peculiar thermal properties due to its unique phase space structure.
We find that the equation of state parameter $\omega_\U$, the ratio
of pressure to energy density, is given by $1/(2d_\U +1)$ providing
a new form of energy in our universe. In an expanding universe, the
unparticle energy density $\rho_\U(T)$ evolves dramatically
differently from that for photons. For $d_\U >1$, even if
$\rho_\U(T_{\rm D})$ at a high decoupling temperature $T_{\rm D}$ is
very small, it is possible to have a large relic density
$\rho_\U(T^0_\gamma)$ at present photon temperature $T^0_\gamma$,
large enough to play the role of dark matter. We calculate $T_{\rm
D}$ and $\rho_\U(T^0_\gamma)$ using photon-unparticle interactions
for illustration.
} 
\maketitle
In cosmology the equation of state (EoS) parameter $\omega$, the
ratio of pressure ($p$) to energy density ($\rho$), for a species of
energy carrier, plays a crucial role in determining the properties
of the expanding universe~\cite{book:cosmo}. It determines the
energy density at a given temperature since $\rho$ evolves with the
Friedmann-Robertson-Walker (FRW) metric scale factor $R$ as
$R^{-3(1+\omega)}$. It also fixes the rate of deceleration since the
deceleration parameter is proportional to $ (1+3\omega)\rho$. For
example, while cold dark matter (CDM) with $\omega_{\rm M} =0$
provides a stronger gravitational attraction than photon whose
$\omega_\gamma = 1/3$, quintessence with $\omega_{\rm Q} <-1/3$ and
cosmological constant $\Lambda$ with $\omega_\Lambda = -1$
accelerate the expansion of our universe. It is an important task
for modern cosmology to determine various relic energy densities and
their EoS parameters~\cite{Seljak:2006bg}. And this has become even
urgent due to the recent discovery in precision cosmological
observations that the majority of the energy budget in our universe
is carried by dark matter and dark energy instead of ordinary matter
\cite{Sadoulet:2007pk}. What is the nature of this `dark side' of
the universe? And is there any alternative to dark matter besides
the often invoked weakly interacting massive particles in
particular?

In this work we demonstrate a novel kind of new energy from
unparticles whose EoS parameter $\omega_\U$ lies between CDM and
photon with $\omega_\U$ equal to $1/(2d_\U +1)$.  It might be dubbed
{\em unmatter}, to be distinguished from CDM and ordinary matter.
Then we investigate some of its impacts on cosmology and
astrophysics. In an expanding universe, the behavior of unparticle
energy density $\rho_\U(T)$ is dramatically different than that for
photons. For $d_\U >1$, even if the density $\rho_\U(T_{\rm D})$ at
a high decoupling temperature $T_{\rm D}$ is very small, it is
possible to have a large relic density $\rho_\U(T^0_\gamma)$ at
present photon temperature $T^0_\gamma$, large enough to play the
role of dark matter.

The concept of unparticle \cite{Georgi:2007ek} stems from the
observation that certain high energy theory with a nontrivial
infrared fixed-point at some scale $\Lambda_{\U}$ may develop a
scale-invariant degree of freedom below the scale, named unparticle.
The notion of mass does not apply to such an identity; instead, its
kinematics is mainly determined by its scaling dimension $d_{\U}$
under scale transformations. The unparticle must interact with
particles, however feebly, to be physically relevant; and the
interaction can be well described in effective field theory (EFT).
There has been a burst of activities since the seminal work of
Georgi~\cite{Georgi:2007ek}, on various aspects of unparticle
physics from precision tests and collider physics effects
\cite{Georgi:2007si,Cheung:2007zza} to theoretical issues
\cite{Stephanov:2007ry,Fox:2007sy} and cosmological and
astrophysical implications~\cite{Davoudiasl:2007jr,Freitas:2007ip},
to mention a few {\cite{apology}. In a glut of unparticle
phenomenological studies, either unparticles are treated at zero
temperature as occurring in ordinary particle physics processes, or
the naive arguments of conformal invariance are invoked for the
unparticle EoS with the massless photon as an analogue in mind. In
this work we work out thermodynamics of unparticles directly from
their basic properties, which turns out to be generally different
from that of photons.

The thermodynamics of a gas of bosonic particles with mass $\mu$ is
determined by the partition function:
\begin{eqnarray}
\ln Z(\mu^2)&=&-g_sV\int\frac{d^4p}{(2\pi)^4}2\pi
2p^0\theta(p^0)\nonumber\\
&&\times\delta(p^2-\mu^2) \ln(1-e^{-p^0\beta}),
\end{eqnarray}
where $V,~\beta=T^{-1}$ are the volume and inverse temperature in
natural units respectively, and $g_s$ accounts for degrees of
freedom like spin. The density of states in four-momentum space is
proportional to the $\delta$ function due to the dispersion relation
for particles. There is no such a constraint in the case of
unparticles, whose density of states is dictated by the scaling
dimension $d_\U$ of the corresponding field to be proportional to
\cite{Georgi:2007ek}:
\begin{eqnarray}
\frac{d^4p}{(2\pi)^4} \theta(p^0)\theta(p^2)(p^2)^{d_\U-2}.
\end{eqnarray}
Nevertheless, we can interpret it in terms of a continuous
collection of particles with the help of a spectral function
$\varrho(\mu^2)\propto\theta(\mu^2)(\mu^2)^{d_\U-2}$
~\cite{Georgi:2007si}:
\begin{eqnarray}
2\pi\theta(p^0)\delta(p^2-\mu^2)
\frac{d^4p}{(2\pi)^4}\varrho(\mu^2)d\mu^2
\end{eqnarray}
In this construction, compared to the case of particles of a
definite mass, $\mu^2$ serves as a new quantum number to be summed
over with the weight $\varrho(\mu^2)$.

To write down the partition function for unparticles, we have to
normalize $\varrho$ correctly. Since unparticles exist only below
the scale $\Lambda_\U$, the spectrum must terminate
there~{\cite{break}}. Beyond the scale, unparticles can be resolved
and are no more the suitable degrees of freedom to cope with. This
also implies that we should require $\beta\LamU> 1$ for
self-consistency. We thus find the normalized spectrum,
\begin{eqnarray}
\varrho(\mu^2)=(\dU-1)\LamU^{2(1-\dU)}\theta(\mu^2)(\mu^2)^{\dU-2},
\end{eqnarray}
which has the correct limit $\delta(\mu^2)$ as $\dU\to 1^+$. Note
that integrability at the lower end of $\mu^2$ requires $\dU\geq 1$.
The partition function for unparticles is
\begin{eqnarray}
\ln Z&=&\int_0^{\LamU^2}d\mu^2\varrho(\mu^2)\ln Z(\mu^2)\nonumber\\
&=&-\frac{g_sV(\dU-1)}{4\pi^2\beta^3(\beta\LamU)^{2(\dU-1)}}
\int_0^{(\beta\LamU)^2}dy~y^{\dU-2}\nonumber\\
&\times&\int_y^{\infty}dx~\sqrt{x-y}\ln(1-e^{-\sqrt{x}})
\end{eqnarray}
For $\beta\LamU> 1$, the above integrals factorize to good precision
due to the exponential:
\begin{eqnarray}
\ln Z&=&\frac{g_sV(\dU-1)~2B(3/2,\dU-1)}
{4\pi^2\beta^3(\beta\LamU)^{2(\dU-1)}(2\dU+1)}\nonumber\\
&\times&\Gamma(2\dU+2)\zeta(2\dU+2),
\end{eqnarray}
where $\Gamma,~B,~\zeta$ are standard functions and integration by
parts has been used for $2\dU+1>0$. Using the definition of $B$
function, the apparent singularity at $\dU=1$ can be removed
explicitly:
\begin{eqnarray}
\ln Z&=&\frac{g_sV}{\beta^3(\beta\LamU)^{2(\dU-1)}}
\frac{\C(\dU)}{4\pi^2},
\end{eqnarray}
with $\C(\dU)=B(3/2,\dU)\Gamma(2\dU+2)\zeta(2\dU+2)$. It is now
straightforward to work out the quantities:
\begin{eqnarray}
p_\U&=&g_sT^4\left({T\over \LamU }\right )^{2(\dU-1)}
\frac{\C(\dU)}{4\pi^2},\nonumber\\
\rho_\U&=&(2\dU+1)g_sT^4\left({T\over\LamU}\right)^{2(\dU-1)}
\frac{\C(\dU)}{4\pi^2}.%
\label{prho}
\end{eqnarray}
Again the case of massless particles is recovered correctly by
setting $\dU=1$ and $\C(1)=2\pi^4/45$. The above results imply the
following EoS parameter for unparticles:
\begin{eqnarray}
\omega_\U=(2\dU+1)^{-1}. %
\label{eos}
\end{eqnarray}
The results for fermionic unparticles can be obtained by replacing
$\C(d_\U)$ by $(1-2^{-(2d_\U+1)})\C(d_\U)$.

It is clear that $\omega_\U$ is very different from that for photons
or CDM, and generically lies in between for $\dU>1$. This is in
contrast to the naive expectation based on conformal theory
arguments and the massless photon analogue. This arises essentially
from the fact that unparticles exist only below a finite energy
scale $\LamU$ as reflected in the spectral function $\varrho(\mu^2)$
while a conventional conformal theory is not characterized by such a
scale. If the limit $\LamU\to\infty$ were naively taken, which means
there would be no unparticles in the infrared, $\rho_\U$ would
vanish trivially. This is indeed not the case interested in here.
The factor $\LamU^{2(1-\dU)}$ in $p_\U,~\rho_\U$} acts as an
effective parameter in the low temperature theory, and the presence
of $\LamU$ reflects its connection to the underlying theory that
produces the unparticle. This connection between low and high energy
theories is completely expected, as for instance, in thermodynamics
of solids viewed from atomic physics.

The ensemble of unparticles thus provides a new form of energy
density in our universe, which will have important repercussions for
cosmology. We now study their implications in our expanding universe
by concentrating on their contribution to the energy density in the
universe. The unparticle energy density at present is determined by
its initial value at the decoupling temperature $T_{\rm D}$ where
unparticles drop out of the thermal equilibrium with standard model
(SM) particles, and its evolution thereafter which is closely
related to the EoS parameter.

In an FRW expanding universe, the energy density $\rho(T)$ (or
$\rho(R)$) of a species after decoupling from equilibrium is given
by
\begin{eqnarray}
\rho(R)=\rho(R_{\rm D})\left({R_{\rm D}\over
R}\right)^{3(1+\omega)}\;,
\end{eqnarray}
where $R_{\rm D}$ is the scale factor of the expanding universe at
decoupling. From now on, we will interchange the notations $\rho(T)$
and $\rho(R)$ freely. Since photon expansion follows $R_{\rm D}/R =
T_\gamma/T_{\rm D}$, we have
\begin{eqnarray}
&&\rho_\U(T_\gamma)=\rho_\U(T_{\rm D})
\left({T_\gamma\over T_{\rm D}}\right )^{3(1+\omega_\U)}\;,\nonumber\\
&&\rho_\gamma(T_\gamma) = \rho_\gamma(T_{\rm D})%
\left({T_\gamma\over T_{\rm D}}\right )^4\;,%
\label{rho_evolution}
\end{eqnarray}
where $\rho_{\gamma},~T_{\gamma}$ are the quantities for photons.
For $d_\U > 1$, the unparticle energy density decreases more slowly
than the photon's as the universe cools down.

\begin{figure}
\resizebox{0.48\textwidth}{!}{%
\includegraphics[width=8cm]{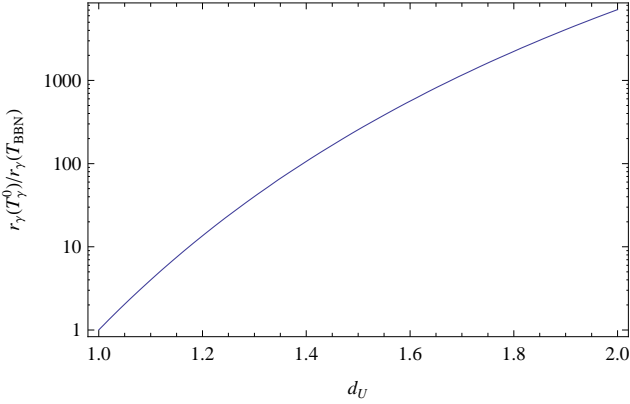}
} %
\caption{The double ratio $r_\gamma(T^0_\gamma)/r_\gamma(T_{\rm
BBN})$ as a function of $d_\U$.} %
\label{deppfig}
\end{figure}

If unparticle is always in thermal equilibrium with photon, its
energy density drops faster than photon when temperature goes down.
However, after unparticle freezes out of equilibrium, the situation
is different. The ratios of the energy densities, $r_\gamma(T)
=\rho_\U(T) /\rho_\gamma(T)$, at two temperatures $T_1$ and $T_2$
are related by
\begin{equation}
r_\gamma(T_2)= r_\gamma(T_1)
\left( \frac{T_1}{T_2}\right)^{\frac{2d_\U-2}{2d_\U+1}}\;.%
\label{depp}
\end{equation}
A dramatic consequence of this is that even if the unparticle
density is small compared with photon's at a high temperature
$T_{\rm D}$, it may become larger or even comparable to the critical
density at a lower temperature. For illustration, we show in Fig.
\ref{deppfig} how the ratio %
$r_\gamma(T_0^\gamma)$ at the present photon temperature $T^0_\gamma
= 2.725\pm 0.002$ K~\cite{PDG} changes with $d_\U$ for a given
$r_\gamma(T_{\rm BBN})$ at the Big-Bang-Nucleosynthesis (BBN)
temperature, $T_{\rm BBN}=1$ MeV, where unparticle and photon are
assumed to have decoupled. We see that the double ratio
$r_\gamma(T^0_\gamma)/r_\gamma(T_{\rm BBN})$ is always larger than
one for $d_\U > 1$.

The above property opens the possibility for unparticle to play an
important role as dark `matter'. Such dark matter, or better named,
unmatter, is different than the usual one. It provides gravitational
attraction, but with EoS deviating from $0$. It would be interesting
to see whether such a picture fits in a global analysis of various
cosmological data. This is however beyond the scope of this work.

The temperature $T_{\rm D}$ depends on unparticle-particle
interactions. In EFT below $\Lambda_\U$, there could be many
possible interactions between unparticles and SM particles even if
unparticles are singlets under the SM gauge
group~\cite{Chen:2007qr}. A practical study with a global fitting
should make a survey of all such interactions and those induced by
thermal effects. For the purpose of illustration here, we consider
below the unparticle-photon interactions:
\begin{eqnarray}
{\cal L} = \lambda\Lambda_\U^{-d_\U} F^{\mu\nu} F_{\mu\nu}\U +
\tilde\lambda \Lambda_\U^{-d_\U} \tilde F^{\mu\nu}F_{\mu\nu}\U\;,
\end{eqnarray}
where $F,~\tilde{F}$ are respectively the electromagnetic field
tensor and its dual, and the coefficients $\lambda,~\tilde{\lambda}$
can be expressed in terms of the standard ones in
Ref.~\cite{Chen:2007qr}. We will treat the two interactions one by
one.

The above interactions can bring photons and unparticles into
equilibrium. Taking the $\lambda$ term as an example, the cross
section for $\gamma \gamma \to \U$ is
\begin{eqnarray}
\sigma(s)={1\over 4}\lambda^2\left( {s\over
\Lambda^2_\U}\right)^{d_\U} {1\over s} A_{d_\U}\;,
\end{eqnarray}
where
\begin{eqnarray}
A_{d_\U}& = &\frac{16\pi^{5/2}\Gamma(d_\U+1/2)}
{(2\pi)^{2d_\U}\Gamma(d_\U-1)\Gamma(2d_\U)}
\end{eqnarray}
is a normalization factor for the unparticle density of states
suggested in Ref.~\cite{Georgi:2007ek}, and the interaction rate is
\begin{eqnarray}
\Gamma \simeq n_\gamma \sigma(s) c &=&\frac
{\zeta(3)A_{d_\U}}{8\pi^2} \lambda^2 T\left (
\frac{2T}{\Lambda_\U}\right )^{2d_\U}\;,
\end{eqnarray}
where $n_{\gamma}$ is the photon number density, and we have used $s
= (2 T)^2$.

This rate is compared with the Hubble parameter $H = 1.66
g_*^{1/2}T^2/m_{\rm Pl}$ in the radiation dominated era to determine
at what temperature unparticles decouple from photons
~\cite{book:cosmo}. Here $g_*$ is the total number of degrees of
freedom at the decoupling temperature and $m_{\rm Pl} = 1.22 \times
10^{19}$ GeV is the Planck mass. When $\Gamma < H$, the unparticles
will decouple from photons. Taking the equal sign, one obtains the
decoupling temperature,
\begin{eqnarray}
T_{\rm D} = {1\over 2}\left ({1.66 g^{1/2}_*\over m_{pl}}
{\Lambda_\U^{2d_\U}\over \lambda^2 A_{d_\U}}{4\pi^2\over
\zeta(3)}\right )^{1/(2d_\U -1)}\;.%
\label{TD}
\end{eqnarray}
Replacing $\lambda$ by $\tilde \lambda$, one obtains the decoupling
temperature due to the $\tilde \lambda$ term. In the following
numerical discussions, we will take $\lambda$ to be non-zero for
illustration. The results will be the same for taking $\tilde
\lambda$ non-zero.

There are experimental constraints on the coupling
$\lambda/\Lambda_\U^{d_\U}$ from
astrophysics~\cite{Davoudiasl:2007jr,Freitas:2007ip}, radiative
positronium decay $\mbox{o-P} \to \gamma \U$~\cite{Chen:2007qr} and
CERN LEP $e^- e^+ \to \gamma \U$ ~\cite{Chen:2007qr}. Among them,
the astrophysical one by energy loss arguments in stars is most
stringent. Using the numbers obtained in Ref.~\cite{Freitas:2007ip}
we can calculate the allowed maximal coupling
$(\lambda/\Lambda_\U^{d_\U})_{\rm max}$ and the corresponding
minimal decoupling temperature $T_{\rm D}^{\rm min}$. The actual
decoupling temperature can of course be higher than this minimal
value. The results are listed in Table~\ref{t1}. It is seen that
$T_{\rm D}^{\rm min}$ can vary in a big range from as large as
$10^7$ GeV to as low as a few 10 GeV depending on the value of
$d_\U$.

\begin{table}
\caption{Upper bound $(\lambda/\Lambda_\U^{d_\U})_{\rm max}$ (in
units of GeV$^{-d_\U}$) and the corresponding $T_{\rm D}^{\rm min}$
(in units of GeV) for various values for $d_\U$. Appropriate $g_*$
has been used for the given energy with SM particles and a scalar
unparticle.} %
\label{t1}
\begin{tabular}{|clll|}
\hline
$d_\U$ & 4/3&5/3&2 \\
\hline
$(\lambda/\Lambda_\U^{d_\U})_{\rm max}$ &
$1.04\times 10^{-14}$&$7.17\times 10^{-13}$&$5.11\times 10^{-11}$\\
$T^{\rm min}_{\rm D}$ &$7.37\times 10^{6}$&$2.70\times 10^{3}$
&$3.68\times 10$\\
\hline
\end{tabular}
\end{table}

\begin{table}
\caption{$\Lambda_\U$ and $r_\gamma(T)=\rho_\U(T)/\rho_\gamma(T)$ as
functions of $\Omega_\U(T^0_\gamma)$. We have used $\rho_{\rm
cr}(T^0_\gamma) = 8.0992h^2\times 10^{-47}$ GeV$^4$ and taken the
central value for $h = 0.73^{+0.04}_{-0.03}$~\cite{PDG}.}
\label{t2}       
\begin{tabular}{|clll|}
\hline
$d_\U = 4/3$&&&\\ 
$\Omega_\U(T^0_\gamma)$ & $1.0$ & $0.161$  &$0.01$  \\
$\Lambda_\U$ (GeV) & $-$ & $7.37\times 10^{6}$ & $4.78\times 10^8$\\
$r_\gamma(T_{\rm D}^{\rm min})$& $-$   &$9.90\times 10^{-1}$
&$6.13\times 10^{-2}$\\
$r_\gamma(T_{\rm BBN})$ & $-$ &$6.22\times 10$ &$3.85$\\ \hline
$d_\U = 5/3$&&&\\ 
$\Omega_\U(T^0_\gamma)$ & $1.0$ &$0.20$ &  $0.01$    \\
$\Lambda_\U$ (GeV) & $1.46\times 10^4$ & $4.87\times10^4$
&$4.61\times 10^5$\\
$r_\gamma(T_{\rm D}^{\rm min}) $ & $2.48\times 10^{-1}$ &$4.97\times
10^{-2}$
&$2.48\times 10^{-3}$\\
$r_\gamma(T_{\rm BBN})$ & $2.35\times 10$ &$4.73$
& $2.36\times 10^{-1}$\\
\hline
$d_\U = 2$&&&\\ 
$\Omega_\U(T^0_\gamma)$& $1.0$ &$0.2$& $0.01$\\
$\Lambda_\U$ (GeV) &$4.31\times 10^2$ &$9.65\times 10^2$
&$4.32\times 10^3$ \\
$r_\gamma(T_{\rm D}^{\rm min})$ &$4.57\times 10^{-2}$
&$9.11\times 10^{-3}$  &$4.55\times 10^{-4}$  \\
$r_\gamma(T_{\rm BBN})$ & $3.06$ &$6.14\times 10^{-1}$
&  $3.05\times 10^{-2}$\\
\hline
\end{tabular}
\end{table}

In order that the relic density of unparticles is not too large, say
as large as the critical energy density ($\rho_{\rm cr}$) which
would over close the universe, for a given $T_{\rm D}$ one has to
choose a big enough $\Lambda_\U$ besides the requirement
$\Lambda_\U>T_{\rm D}$. This provides a way to constrain the scale
$\Lambda_\U$ directly. We illustrate our results in Table \ref{t2}
for several representative values of the ratio of energy densities,
$\Omega_\U (T^0_\gamma) = \rho_\U(T^0_\gamma)/\rho_{\rm
cr}(T^0_\gamma)$, at $T^0_\gamma$. In our analysis, we assume
$T_{\rm D}=T_{\rm D}^{\rm min}$, as shown in Table \ref{t1}. By
equating $\rho_\U(T_{\rm D})$ that is obtained via eq. (\ref{prho})
on the one hand with the one from backward evolution via eq.
(\ref{rho_evolution}) on the other, we can determine $\LamU$ for
each given $d_\U$. Also shown are the values of the ratio
$r_\gamma=\rho_\U/\rho_\gamma$ at $T_{\rm D}^{\rm min}$ and $T_{\rm
BBN}$. Note that we do not assume a value for the dimensionless
coupling $\lambda$; instead, it is fixed by $\LamU$ and $T_{\rm
D}=T_{\rm D}^{\rm min}$ via eq. (\ref{TD}).

For $d_\U=4/3$, we find that it is not possible to saturate the
critical density, nor the dark matter density $\Omega_{\rm DM} =
0.2$~\cite{PDG}. With the constraint that $T_{\rm D} < \tilde
\Lambda_\U$, the largest $\Omega_\U(T^0_\gamma)$ is 0.16 which
occurs at $T_{\rm D}=\tilde \Lambda_\U$. This, of course, still
leaves enough room for unparticle to play a significant role as dark
matter. For $d_\U=5/3$ and $2$, we see that the present unparticle
relic density can easily saturate the critical density and dark
matter density. In all cases, $\rho_\U(T_{\rm D})$ is smaller (in
most cases much smaller) than $\rho_\gamma(T_{\rm D})$. Requiring
the present relic of unmatter to be less than these densities one
obtains conservative lower bounds on $\tilde \Lambda_\U$ for given
$T_{\rm D}=T_{\rm D}^{\rm min}$. For small $d_\U$, the scale $\tilde
\Lambda_\U$ is constrained to be very large, making low energy
search for unparticle effects difficult. But for large $d_\U$ (close
to 2), the scale can still be as low as a few hundred GeV which may
be reached at LHC and ILC colliders.

The standard BBN theory explains data well. It is therefore
important to make sure that at $T_{\rm BBN}$ unparticles do not
cause problems. A simple criterion is to require that at this
temperature the unparticle energy density be less than the photon's.
With this restriction, it is interesting to see whether one can
still have large relic unmatter at present. We find this is indeed
possible. Although there are many cases shown in Table~\ref{t2}
where $r_\gamma(T_{\rm BBN})$ is larger than one, circumstances with
sizable $\Omega_\U(T^0_\gamma)$ but small $r_\gamma(T_{\rm BBN})$
also appear at large $d_\U$. This can easily be understood from eq.
(\ref{depp}) and from Fig.\ref{deppfig}. For $d_\U >1$, a small
$r_\gamma(T_{\rm BBN})$ can result in a sizable
$\Omega_\U(T^0_\gamma)$. A universe dominated by unparticle between
the BBN era and the matter or dark energy dominated universe is
possible.

In the above discussions, the interactions of unparticles with
photons lead to an interaction rate $\Gamma \sim T^{2d_\U+1}$ which
brings unparticles and SM particles into equilibrium at a high
temperature, and they decouple at a lower temperature if the
unparticle dimension $d_\U$ is larger than 1. There are many
possible ways unparticles can interact with SM particles , but not
all interactions will have the same properties as far as thermal
equilibrium is concerned. For example, we find that all of the
operators involving SM fermions listed in Ref.\cite{Chen:2007qr}
will result in an interaction rate proportional to $T^{2d_\U-1}$. If
unparticles and SM fermions are required to be in thermal
equilibrium at a high temperature and then decouple at a lower one,
$d_\U$ must be larger than $3/2$. On the contrary, if $d_\U$ is less
than 3/2, then unparticles and SM fermions will not be in thermal
equilibrium at a high temperature in the first place, but will be at
a lower temperature till the epoch of matter dominated universe.
Since when in thermal equilibrium, the unparticle density dilutes
faster than SM particles, its relic density today will be negligibly
small if the equilibrium sets in before or just after BBN with the
unparticle relic density not larger than photon density. This is an
interesting scenario to study, which may lead to sensitive
information about the unparticle scaling dimension.

There is much to be explored for the roles that thermal unparticles
can play in our universe. It is important to analyze available
cosmological and astrophysical data for a global fit with unparticle
energy density integrated. We leave this for a detailed future
study.

To summarize, we have studied for the first time the thermal
properties of unparticles. Due to its peculiar phase space structure
we found that the EoS parameter $\omega_\U$ is given by $1/(2d_\U
+1)$, providing a new form of energy in our universe. In an
expanding universe, the behavior of unparticle energy density
$\rho_\U(T)$ is dramatically different than that for photons. For
$d_\U >1$, even if its value at a high decoupling temperature
$T_{\rm D}$ is very small, it could evolve into a sizable relic
density $\rho_\U(T^0_\gamma)$ at present, large enough to play the
role of dark matter. We have exemplified this with photon-unparticle
interactions, and found that it is indeed feasible to obtain a large
relic energy density of unparticles with the most stringent
constraints saturated.

\vskip .2cm %
\noindent %
{\bf Acknowledgments} This was supported in part by the grants
NSC95-2112-M-002-038-MY3, NCTS (NSC96-2119-M-002-001), NCET-06-0211,
and NSFC-10775074.



\begin{thebibliography}{}
\bibitem{break}We assume scale invariance all the way up to $\Lambda_\U$
throughout the paper though it could potentially be broken by
couplings to standard model fields at lower energies.
\end{thebibliography}
\end{document}